\documentclass[
    ,final            
  ]
  {aipproc}
\layoutstyle{8x11double}
\usepackage{clshan-math,clshan-dm-dd}
\newcommand{\beenr}{\begin{enumerate}}
\newcommand{\eeenr}{\end{enumerate}}
\begin{document}
\title{The {\tt AMIDAS} Website: An Online Tool for \\
       Direct Dark Matter Detection Experiments}
\author{Chung-Lin Shan}
       {address={School of Physics and Astronomy, Seoul National University,
                 Seoul 151-747, Republic of Korea \\
                 E-mail: {\tt cshan@hep1.snu.ac.kr}}}
\begin{abstract}
 Following our long--term work on development of
 model--independent data analysis methods
 for reconstructing the one--dimensional
 velocity distribution function of halo WIMPs
 as well as for determining their mass and couplings on nucleons
 by using data
 from direct Dark Matter detection experiments directly,
 we combined the simulation programs to a compact system:
 {\tt AMIDAS} (A Model--Independent Data Analysis System).
 For users' convenience an online system
 has also been established at the same time.
 {\tt AMIDAS} has the ability to do
 full Monte Carlo simulations,
 faster theoretical estimations,
 as well as to analyze (real) data sets
 recorded in direct detection experiments
 {\em without} modifying the source code.
 In this article,
 I give an overview of functions of the {\tt AMIDAS} code
 based on the use of its website.
\end{abstract}
\maketitle
%
%
\section{Introduction}
 Weakly Interacting Massive Particles (WIMPs) $\chi$
 arising in several extensions of
 the Standard Model of electroweak interactions
 are one of the leading candidates for Dark Matter.
 Currently,
 direct searches for different candidates for WIMP Dark Matter
 based on measuring recoil energy
 deposited by elastic scattering of ambient WIMPs on the target nuclei
 \cite{Smith90, Lewin96}
 are one of the promising methods
 for understanding the nature of Dark Matter particles,
 identifying them among new particles
 produced hopefully in the near future at colliders,
 as well as reconstructing the (sub)structure of our Galactic halo.

 However,
 for conventional data analyses
 used in direct detection experiments
 one needs assumptions not only about
 the Galactic halo from astrophysics
 but also about the WIMP properties
 from particle physics \cite{SUSYDM96}.
 Therefore,
 since a few years ago
 we started to develop new methods
 for analyzing data, i.e., measured recoil energies,
 from (future) direct detection experiments
 as model--independently as possible.
 Up to now we could in principle
 reconstruct the (moments of the) one--dimensional
 velocity distribution function of halo WIMPs \cite{DMDDf1v},
 as well as determine the WIMP mass \cite{DMDDmchi}
 and (ratios of) their couplings on nucleons
 \cite{DMDDfp2-IDM2008, DMDDidentification-Dark2009}.

 Following the development of
 these model--independent data analysis procedures,
 we combined the programs for simulations to a compact system:
 {\tt AMIDAS} (A Model--Independent Data Analysis System).
 For users' convenience and
 under the collaboration with the ILIAS Project \cite{ILIAS},
 an online system \cite{AMIDAS-web}
 has also been established at the same time.

 In this article,
 I give an overview of functions of the {\tt AMIDAS} code
 based on the use of its website.
 In Sec.~2
 I will describe the {\tt AMIDAS}' functions and
 different working modes for these functions.
 In Sec.~3
 I will talk about the options of the input parameters
 for simulations which users can modify.
 In Sec.~4
 the use of the {\tt AMIDAS} website
 for analyzing user--uploaded data sets
 will be described.
 I will conclude and
 give some future prospects of the {\tt AMIDAS} code and website
 in Sec.~5.
\section{Functions, modes, targets}
\subsection{{\tt AMIDAS}' Functions}
 Based on our works on
 the model--independent data analysis methods
 for extracting the nature of halo WIMPs
 \cite{DMDDf1v, DMDDmchi, DMDDfp2-IDM2008, DMDDidentification-Dark2009},
 {\tt AMIDAS} has so far the following functions:
\beenr
\item
 reconstruction of the one--dimensional
 velocity distribution function of halo WIMPs;
\item
 determination of the WIMP mass;
\item
 determinations of ratios of
 different WIMP--nucleon couplings/cross sections;
\item
 estimation of the spin--independent (SI) WIMP--proton coupling.
\end{enumerate}
\subsection{Reconstruction modes}
 For reconstructing
 the one--dimensional WIMP velocity distribution
 and estimating the SI WIMP--proton coupling,
 one needs the WIMP mass $\mchi$
 as an input parameter
 \cite{DMDDf1v, DMDDfp2-IDM2008}.
 This information could be obtained
 either from e.g., collider experiments,
 or from two direct detection experiments \cite{DMDDmchi}.
 To allow these two cases,
 {\tt AMIDAS} has three options of the input WIMP mass
 for the reconstruction mode:
\beenr
\item
 with only an input WIMP mass
 from other/collider experiments;
\item
 with only a reconstructed WIMP mass
 from other direct detection experiments;
\item
 with both of them.
\end{enumerate}
 In addition,
 {\tt AMIDAS} offers two modes for determining the WIMP mass
 \cite{DMDDmchi}:
\beenr
\item
 only the combined result
 from the estimators for different moments;
\item
 both the combined result from and each of
 the estimators for different moments.
\end{enumerate}
\subsection{Target(s)}
 So far {\tt AMIDAS} uses
 $\rmXA{Si}{28}$, $\rmXA{Ge}{76}$,
 $\rmXA{Ar}{40}$, and $\rmXA{Xe}{136}$
 for simulations and
 can analyze data sets with these four targets.
 For the determination of the WIMP mass,
 two combinations have been programmed \cite{DMDDmchi}:
\beenr
\item
 $\rmXA{Si}{28}$ + $\rmXA{Ge}{76}$;
\item
 $\rmXA{Ar}{40}$ + $\rmXA{Xe}{136}$.
\end{enumerate}
\subsection{Data type}
 The most important and powerful ability
 of the {\tt AMIDAS} code is that
 this system and its website
 can {\em not only} do simulations with self--generated events
 based on the Monte Carlo method,
 {\em but also} analyze user uploaded data set(s)
 generated by other event generators or
 {\em recorded} in direct Dark Matter detection experiments
 {\em without} modifying the source code.
 Users have thus two choices for the data type:
\beenr
\item
 a simulation (events will be generated by {\tt AMIDAS});
\item
 real (user--uploaded) data.
\end{enumerate}
 A sample file for the uploaded data sets
 can be downloaded from the {\tt AMIDAS} website.
\subsection{Simulation mode}
 All {\tt AMIDAS}' functions can be simulated
 based on the Monte Carlo method.
 Considering the current experimental sensitivity
 and the required executing time for these simulations,
 the {\tt AMIDAS} website offers full Monte Carlo simulations
 with maximal 2,000 experiments and
 maximal 2,500 events on average per one experiment.

 However,
 since the algorithmic procedure
 for the determination of the WIMP mass,
 needed also for the reconstruction of
 the one--dimensional WIMP velocity distribution function
 and for the estimation of the SI WIMP--proton coupling,
 takes a (much) longer time than
 what usual Monte Carlo simulations require,
 {\tt AMIDAS} also offers users faster theoretical estimations
 as an alternative option:
\beenr
\item
 a Monte Carlo simulation;
\item
 a theoretical estimation.
\end{enumerate}
 Here integrals over the theoretical predicted recoil spectrum
 will be used.

 Note that,
 firstly,
 since for these estimations
 the statistical fluctuations
 {\em have not} been taken into account,
 these pure theoretically estimated results,
 especially for cases with (very) few events,
 could be (fairly) different from
 results obtained by more realistic simulations.
 Secondly,
 as the alternative option for the Monte Carlo simulation
 with a much shorter required executing time,
 the total event number used for theoretical estimations
 is fixed%
\footnote{
 The actual event number for
 each Monte Carlo simulated experiment
 is Poisson--distributed around
 the expected value set by users.
}
 and the calculations are limited to be done for only a few times.
 These restrictions could cause sometimes
 unexpected zigzag on the result curves.
\section{Running simulations}
 In this section,
 I describe the options of the input factors
 needed for predicting the recoil spectrum
 used {\em only} for generating events.
 Note that
 some commonly used and standard {\tt AMIDAS}' values
 used for our works presented
 in Refs.~\cite{DMDDf1v, DMDDmchi,
                DMDDfp2-IDM2008, DMDDidentification-Dark2009}
 have been given as default choices,
 but all these parameters can be modified by users.
\subsection{WIMP properties}
 The following information on the WIMP properties
 are required for predicting the recoil spectrum
 and/or analyzing user--uploaded data.
 Note that {\em not} all of them
 are needed for every {\tt AMIDAS}' function.
\beenr
\item
 the input WIMP mass $\mchi$;
\item
 an overall uncertainty on the input WIMP mass $\sigma(\mchi)$;
\item
 the SI WIMP--proton cross section $\sigmapSI$.
\end{enumerate}
\subsection{Astronomical setup}
 {\tt AMIDAS} requires the following astronomical parameters
 for the velocity distribution function of halo WIMPs:
\beenr
\item
 the WIMP density near the Earth $\rho_0$;
\item
 the Sun's orbital velocity in the Galactic frame $v_0$;
\item
 the escape velocity from our Galaxy
 at the position of the Solar system $\vesc$;
\item
 the date on which the Earth's velocity
 relative to the WIMP halo is maximal,
 $t_{\rm p}$;
\item
 the experimental running date $t_{\rm expt}$.
\end{enumerate}
\subsection{Velocity distribution of halo WIMPs}
 So far users have two options
 for the one--dimensional WIMP velocity distribution function
 \cite{DMDDf1v}:
\beenr
\item
 the simple Maxwellian velocity distribution
 $f_{1, \Gau}(v)$;
\item
 the shifted Maxwellian velocity distribution
 $f_{1, \sh}(v)$.
\end{enumerate}
 Note that
 the analytical forms of these velocity distributions
 can be checked on the website,
 once one hovers the curser onto the ``{\tt analytical form}''.
 By clicking it one can also open another page
 and get more detailed information
 and some references
 about the velocity distribution.
\subsection{Nuclear form factor for the SI cross section}
 So far users have four options
 for the nuclear form factor
 for the SI WMP--nucleus cross section:
\beenr
\item
 the exponential form factor $F_{\rm ex}(Q)$
 \cite{Ahlen87, Freese88, SUSYDM96};
\item
 the Woods-Saxon form factor $F_{\rm WS}(Q)$
 \cite{Engel91, SUSYDM96};
\item
 the Woods-Saxon form factor
 with a modified nuclear radius $F_{\rm WS, Eder}(Q)$
 \cite{Eder68, Lewin96};
\item
 the Helm form factor $F_{\rm Helm}(Q)$
 \cite{Helm56, Lewin96}.
\end{enumerate}
 As for the velocity distribution function,
 the analytical forms of these form factors
 can be checked on the website,
 by hovering the curser onto the ``{\tt analytical form}''
 or by clicking it to open another page
 for more detailed information and references.
\subsection{Nuclear form factor for the SD cross section}
 For the nuclear form factor
 for the spin--dependent (SD) WIMP cross section,
 due to its dependence on the SD WIMP--nucleon couplings
 as well as on the individual spin structure of target nuclei,
 {\tt AMIDAS} offers so far only one analytic form
 for the SD cross section,
 namely, 
\beenr
\item
 the thin-shell form factor, $F_{\rm TS}(Q)$
 \cite{Lewin96, Klapdor05}.
\end{enumerate}
\subsection{Experimental setup}
 Finally,
 one needs to set the following experimental information:
\beenr
\item
 the minimal and maximal cut--off energies, $\Qmin$ and $\Qmax$;
\item
 the width of the first $Q-$bin, $b_1$;
\item
 the (expected) total event number
 between $\Qmin$ and $\Qmax$, $N_{\rm tot}$;
\item
 the number of simulated experiments or
 uploaded data sets;
\item
 the number of $Q-$bins between $\Qmin$ and $\Qmax$.
\end{enumerate}
 Note that,
 as for the WIMP velocity distribution function
 and the elastic nuclear form factors,
 users can hover the curser onto each notation
 in the setup tables for checking its definition.
\subsection{Running simulations}
 After giving all the required information
 for the aimed simulation,
 users have one more chance to check their choices,
 modify some of them,
 and then resubmit the whole setup.
 In case that any required datum is missed,
 this omission will be detected
 automatically after the (re)submission;
 users will be reminded of that
 with a {\em red} block around the table.
 Note that
 {\em all data} in this table
 will be {\em reset} to the default values
 and should therefore be {\em checked again} and
 modified eventually to the users' own choices.

 Once all the required data have been checked,
 users have only to click
 the ``{\tt Simulation start}'' button
 and wait for the simulation results for a few minutes.
\subsection{Output results}
 Simulation results will be given in form(s) of plot(s),
 and/or eventually table(s).
 In order to let users understand the output results
 more conveniently and clearly,
 each output plot or table
 will be accompanied with a short caption.
 On the other hand,
 for users' need of self--producing
 results with different kinds of presentation,
 the original file of output results
 with users' personal simulation setup
 will also be given and downloadable from the website.
 Remind that
 it would be very grateful that
 a credit of the {\tt AMIDAS} program and website
 could be given for using the output results.
\section{Analyzing (real) data}
 The most useful and powerful function of the {\tt AMIDAS} website
 is the ability for analyzing
 user--uploaded data set(s) directly.
\subsection{Preparing data set(s)}
 As mentioned above,
 on the {\tt AMIDAS} website users can find and download
 a sample file for the uploaded data sets.
 Note that,
 for comments
 a ``0'' (zero) {\em has to be used} at the beginning;
 and all words in the comment lines
 must be connected by ``\_'' (underscores).
 For instance,
{\footnotesize
\begin{verbatim}
  0   sigma_[chi,_p]^SI_=_1e-8_pb
  0   m_[chi]_=_50_GeV

     1 dataset, 43 events, 12817.27 kg-day:
        1      1      5.25   keV;
        1      2      5.37   keV;
               :

  0   m_[chi]_=_100_GeV

     2 dataset, 53 events, 15322.9  kg-day:
        2      1      6.16   keV;
        2      2      4.25   keV;
               :
\end{verbatim}
}
\noindent
 Note also that
 it is {\em unnecessary}
 to output generated/recorded recoil energies
 in the ascending or descending order
 in your uploaded data file(s).
 {\tt AMIDAS} will order the events in each data set
 after reading them.
\subsection{Uploading data file(s)}
 Users can upload their data file(s) as usual.
 Note only that
 the maximal size of each uploaded file is {\em 2 MB}.
\subsection{Analyzing uploaded data}
 As for simulations,
 after giving all the required information
 for the aimed analysis,
 users have one more chance to check their choices and
 the {\em original} name(s) of their data file(s),
 modify some of them and/or
 replace the uploaded data file(s),
 and then resubmit the whole setup.
 In case that any required datum or {\em data file} is missed,
 this omission will be detected
 automatically after the (re)submission;
 users will be reminded of that
 with a {\em red} block around the table.
 Note that,
 while {\em all data} in this table
 will be {\em reset} to the default values
 and should therefore be checked again and
 modified eventually to the users' own choices,
 {\em only} the {\em missed} data file(s)
 and the {\em replacement(s)} of the uploaded file(s)
 will be required to upload.

 Once all the required data and uploaded data file(s)
 have been checked,
 users have only to click
 the ``{\tt Data analysis start}'' button
 and wait for the analyzed results for a few minutes.
\section{Summary}
 In this article,
 I introduced a new simulation/data analysis code and its website
 for direct Dark Matter detection experiments.
 So far users have only a few options
 for the WIMP velocity distribution function
 as well as for the nuclear form factors,
 as a planned improvement
 user--defined velocity distribution and
 form factor(s) for their own target(s)
 should be able to read from an uploaded plain text file
 in the future.
 Moreover,
 the choice(s) of target(s) will also be released
 for (at least) most of the currently running and projected experiments.
%
%
%
%
%
%
\begin{theacknowledgments}
 The author would like to thank the ILIAS Project and
 the Physikalisches Institut der Universit\"at T\"ubingen
 for kindly providing the opportunity of the collaboration
 and the technical support of the {\tt AMIDAS} website.
 This work was partially supported
 by the BK21 Frontier Physics Research Division under project
 no.~BA06A1102 of Korea Research Foundation.
\end{theacknowledgments}
%
%
%
\bibliographystyle{aipproc}   
%
%
\bibliography{sample}
%
%
\IfFileExists{\jobname.bbl}{}
 {\typeout{}
  \typeout{******************************************}
  \typeout{** Please run "bibtex \jobname" to optain}
  \typeout{** the bibliography and then re-run LaTeX}
  \typeout{** twice to fix the references!}
  \typeout{******************************************}
  \typeout{}
 }
%
%
%

%
%
\end{document}